\documentclass[aip,apl,reprint]{revtex4-2}

\usepackage{amsmath,amssymb,graphicx,bm}
\usepackage{graphicx}
\usepackage{subcaption}

\begin{document}

\title{The Study of Thermal Fluctuations in Microwave and Mechanical Resonators}

\author{Michael T Hatzon}
\affiliation{Quantum Technologies and Dark Matter Research Lab, Department of Physics, University of Western Australia, Crawley, Western Australia 6009, Australia}
\author{Eugene N Ivanov}
\affiliation{Quantum Technologies and Dark Matter Research Lab, Department of Physics, University of Western Australia, Crawley, Western Australia 6009, Australia}
\author{Aaron Quiskamp}
\affiliation{Quantum Technologies and Dark Matter Research Lab, Department of Physics, University of Western Australia, Crawley, Western Australia 6009, Australia}
\author{Michael E Tobar}
\affiliation{Quantum Technologies and Dark Matter Research Lab, Department of Physics, University of Western Australia, Crawley, Western Australia 6009, Australia}

\date{\today}

\begin{abstract}
We report high-resolution measurements of thermal fluctuations in microwave and mechanical resonators using a dual-channel readout system. The latter comprises a low-noise amplifier, an I/Q-mixer, and a cross-correlator. We discovered that, under certain conditions, the intrinsic fluctuations of the low-noise amplifier, which are common to both channels of the readout system, are averaged out when computing the voltage noise cross-spectrum between the mixer's outputs. The suppression of the amplifier's technical fluctuations significantly improves the contrast of the thermal noise peaks exhibited by the resonators. Thus, for the room-temperature-stabilized 9 GHz sapphire-loaded cavity resonator, we observed more than 16 dB improvement in the thermal noise peak contrast relative to the single-channel measurements. The ability of the dual-channel readout system to discriminate between the broad- and narrow-band fluctuations may benefit the search for dark matter, which relies on the use of cryogenic microwave resonators.
\end{abstract}

\maketitle
High-resolution noise measurements are crucial for testing fundamental physics \cite{mcallister2018cross,wolf2004improved,abbott2016observation,myers2019high}, characterizing electromagnetic oscillators \cite{ivanov2009low,rubiola2008phase}, and developing secure communication links \cite{li2012secure} and Doppler radars \cite{doerry2018radar}. Techniques such as cross-correlation signal processing \cite{walls1988extending,Walls1999MeasurementOF,rubiola1999improved,rubiola2000correlation,ivanov2002interpreting,kronowetter2023quantum}, microwave circuit interferometry \cite{ivanov2002interpreting}, and power recycling \cite{ivanov2002real} have enabled noise measurements with spectral resolution below the standard thermal noise limit \cite{ivanov2009microwave}. Those advances were largely assisted by the emergence of low-noise HEMT microwave amplifiers \cite{ivanov2005study}. Last but not least, microwave interferometry played a crucial role in elucidating the nature of excess noise in microwave signals extracted from stable optical references via the femtosecond laser comb technique \cite{baynes2015attosecond,fortier2016optically}.

This work demonstrates that the dual-channel readout system with cross-correlation signal processing enables more accurate characterization of thermal fluctuations exhibited by high-Q resonators compared to what can be achieved with conventional single-channel readout.

\begin{figure}[htbp]
    \centering
    \begin{subfigure}[b]{0.54\linewidth}
        \includegraphics[width=\linewidth]{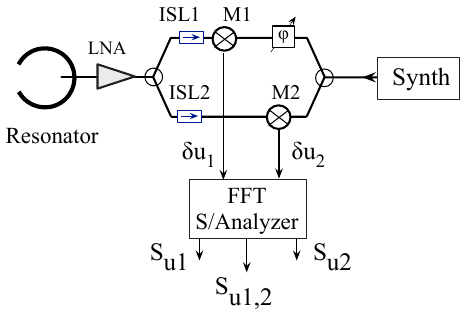}
        \caption{}
        \label{fig:Fig1a}
    \end{subfigure}
    \hfill
    \begin{subfigure}[b]{0.44\linewidth}
        \includegraphics[width=\linewidth]{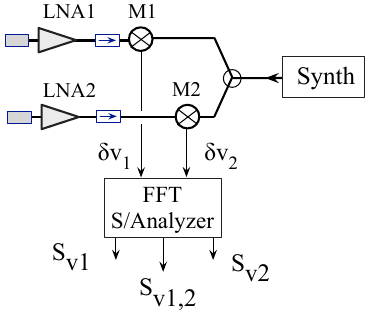}
        \caption{}
        \label{fig:Fig1b}
    \end{subfigure}
    \caption{Dual-channel microwave readout systems.}
    \label{fig:1}
\end{figure}

Fig.~1(a) shows the schematic diagram of the dual-channel readout system we used for the observation of the thermal noise spectra of the microwave resonators. The readout comprises a Low-noise Amplifier (LNA), Wilkinson power dividers, isolators (ISL1, 2), and double-balanced mixers (M1, 2). A microwave frequency synthesizer driving the mixers is sufficiently detuned from the resonator’s eigenfrequency to minimize the mixers' 1/f-noise contribution to the beat note spectra. The Fast Fourier Transform (FFT) spectrum analyser in Fig. 1 computes Power Spectral Densities (PSD) of the individual beat notes $S_{u1}$ and $S_{u2}$, as well as the cross-voltage PSD between the mixer outputs $S_{u1,2}$.

The phase shifter $\varphi$ in the upper arm of the measurement system in Fig.~1 is adjusted to minimize the absolute value of the cross-spectrum $|S_{u1,2}|$. Fig.~2 shows the ratio $|S_{u1,2}|$/$S_{u1}$ as a function of the phase shift $\varphi$. The data in Fig.~2 were obtained at 9 GHz with the LNA input terminated by $50$ $\Omega$, and the number of averages, $N_{avg}$, equal to $2^{14}$. The notch in Fig.~2 has a depth of $\approx 16$ dB. This implies that random processes at the mixer outputs are not entirely independent, as the calculations suggest. Indeed, for two independent noise processes, the depth of the notch (at given number of averages) is expected to be approximately $5log_{10}(N_{avg}) \sim 21$ dB. The latter value was measured using an auxiliary readout system shown in Fig. 1(b). The possible reasons for the residual cross-talk between the mixer outputs include the limited resolution of setting the phase shift $\varphi$ and variations in ambient temperature that affect it.

\begin{figure}
    \centering
    \includegraphics[width=0.9\linewidth]{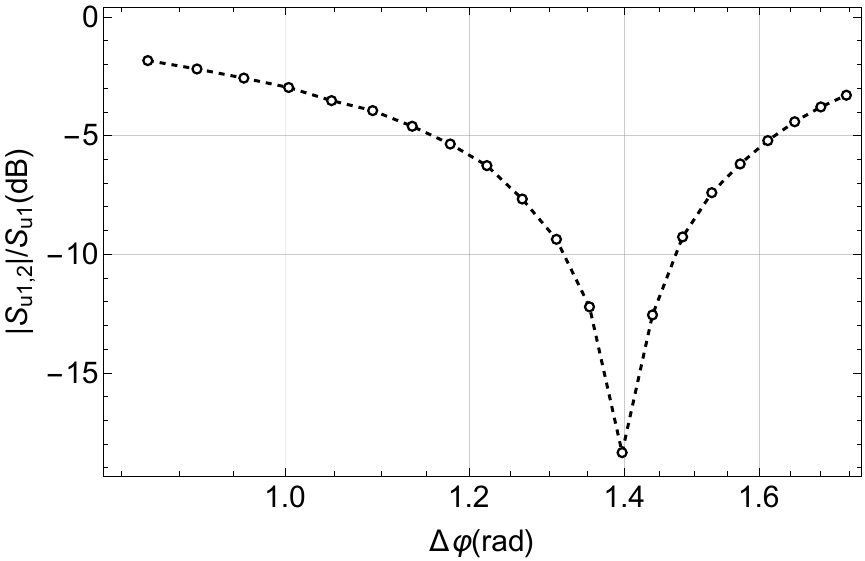}
    \caption{Relative cross-voltage noise power spectral density as a function of the phase shift between two mixer signals.}
    \label{fig:2}
\end{figure}

Having set the phase shift $\varphi$ minimizing the cross-voltage PSD $|S_{u1,2}|$, we proceeded by measuring the thermal noise of a high-Q resonator. The latter was a temperature-stabilized Sapphire Loaded Cavity (SLC) resonator excited in its fundamental whispering gallery TE-mode with azimuthal number $m = 5$. The coupling to the resonator was set close to critical to maximize the emitted thermal noise power. The resonator loaded bandwidth $\Delta f^{(L)}_{res}$ was close to 90 kHz, corresponding to a loaded quality factor Q$_{L} \sim 100$ $000$.

Fig.~3 shows the rms voltage noise spectra $\delta_{u1}=\sqrt{S_{u1}}$ and $\delta_{u2}=\sqrt{S_{u2}}$ (curves 1 and 2) and the rms cross-voltage noise $\delta u_{12}=\sqrt{|S_{u1,2}|}$ (curve 3). The detuning between the resonator and synthesizer was about 600 kHz. The measurement time was approximately half an hour, during which $10^{5}$ averages were taken. Defining the thermal noise peak contrast as a ratio of the peak height to its pedestal, the improvement in the peak contrast exceeded 17 dB when measuring the cross-spectra instead of the single-channel spectra.

\begin{figure}
    \centering
    \includegraphics[width=0.99\linewidth]{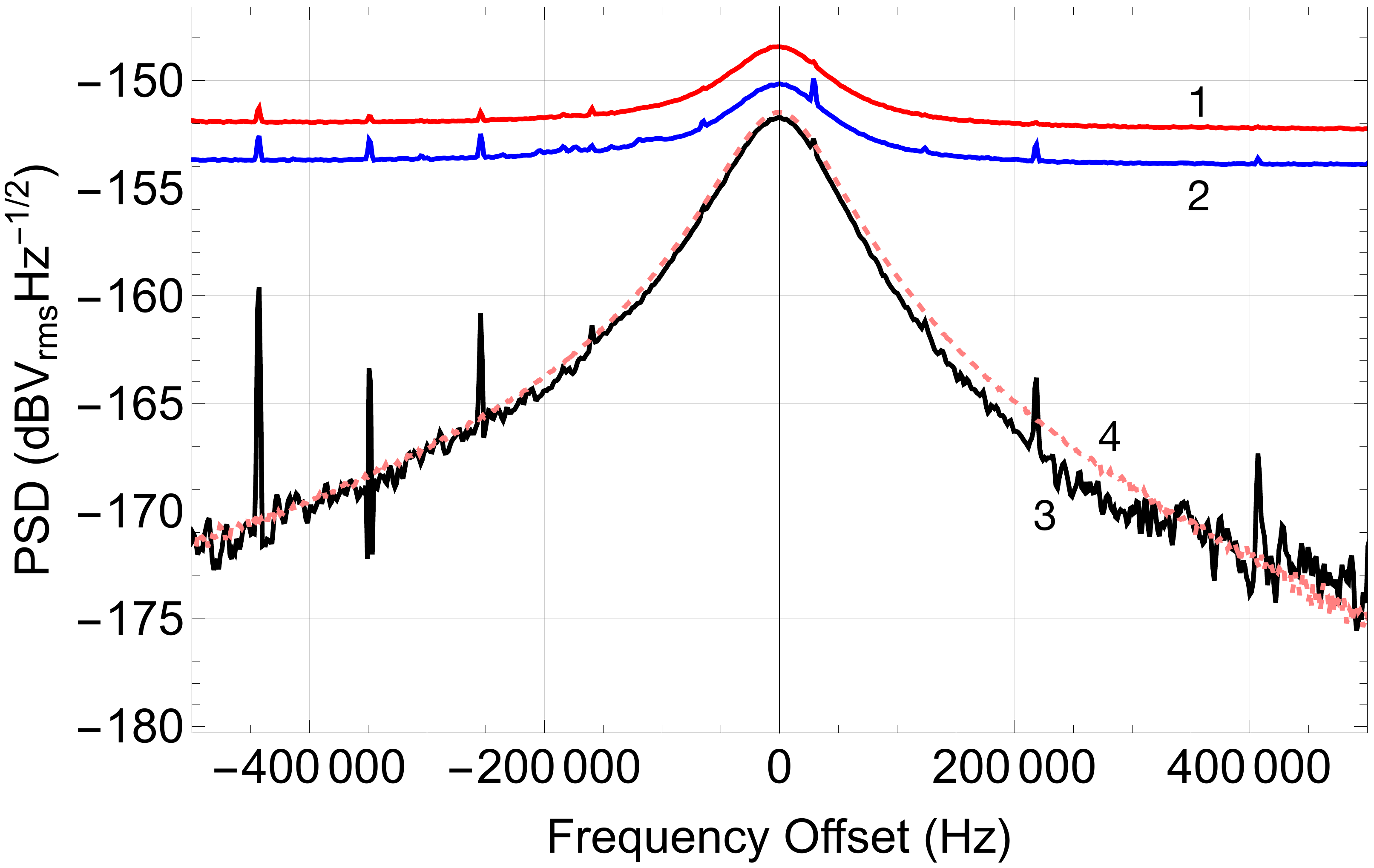}
    \caption{Rms voltage fluctuations at the outputs of the individual channels (curve 1 and 2), and cross-voltage rms noise (curves 3 and 4).}
    \label{fig:3}
\end{figure}

The dashed trace 4 in Fig.~3 is a cross-voltage rms noise computed from the approximate expression (whose derivation will be discussed later)
\begin{equation}
    \delta u_{1,2} = SF \chi_{mix} \sqrt{k_{B} T_{res} (1-|S_{11}|^{2}) \alpha K_{LNA}},
\end{equation}
where $SF$ is the dimensionless scale factor characterizing mixer operation, $\chi_{mix}$ is the mixer power-to-voltage conversion efficiency, $k_{B}$ is the Boltzmann constant, $T_{res}$ is the resonator physical temperature, $S_{11}$ is the resonator reflection coefficient, $\alpha$ is the dimensionless factor characterizing the total loss from the LNA output to the mixer RF input, and $K_{LNA}$ is the LNA power gain. 

For a mixer acting as an ideal multiplier of two microwave signals, calculations show that $SF \sim \sqrt{2}$. Parameter $\chi_{mix}$ was inferred from the measurements of the beat note produced by two frequency-detuned microwave oscillators by changing the power of a signal at the mixer’s RF port. For both mixers, the value of the $\chi_{mix}$ was approximately 7.5 $V/\sqrt{W}$. The SLC temperature, $T_{res}$, was stabilised at 278 K. Finally, a Vector Network Analyzer was used to measure the remaining parameters ($S_{11}$, $\alpha$, and $K_{LNA}$) in Eq.~1. Considering the uncertainties in determining the parameters in Eq.~1, there is a good consistency between the measured and computed noise spectra.

Eq.~1 was derived using the equivalent circuit of the resonator-amplifier assembly shown in Fig.~4. Here, the resonator is presented as a series LCR circuit with the complex impedance $Z_{res}$ magnetically coupled (via the mutual inductance M) to the amplifier input. The resonator coupling coefficient is defined as $\beta = (\omega_{res}  M)^{2}$⁄$(\rho r)$, where $\omega_{res}$ is the resonator angular resonant frequency, $r$ is the resonator loss, and $\rho$ is the amplifier input resistance (also equal to the characteristic impedance of the transmission line connecting the resonator and amplifier).

\begin{figure}
    \centering
    \includegraphics[width=1\linewidth]{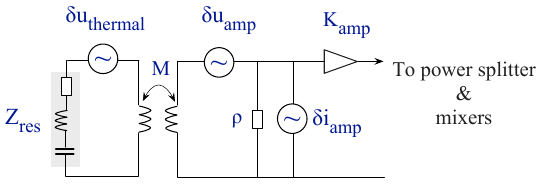}
    \caption{Resonator-Amplifier noise model}
    \label{fig:4}
\end{figure}

The voltage and current noise sources in Fig.~4 are statistically independent, and their contributions to the total voltage noise referred to the amplifier input, $\delta u_{\Sigma}$, can be calculated separately. Thus, the PSD of voltage noise $\delta u_{\Sigma}$ due to the resonator thermal fluctuations can be expressed as 
\begin{equation}
    S^{thermal}_{u}=k_{B} T_{res} \rho \frac{4 \beta}{(1+\beta)^{2}+\zeta^{2}},
\end{equation}
where $\zeta$ is the relative frequency offset from the resonance, i.e., $\zeta = 2(f - f_{res})$/$\Delta f_{res}$. Here, $\Delta f_{res}$ is the intrinsic bandwidth of the resonator.

The technical fluctuations of the LNA are presented by voltage and current sources, $\delta u_{amp}$ and $\delta i_{amp}$, respectively. Their combined contribution to the LNA input voltage noise can be presented as
\begin{equation}
    S^{technical}_{u}=2 k_{B} T_{amp} \rho \frac{1+\beta^{2}+\zeta^{2}}{(1+\beta)^{2}+\zeta^{2}}
\end{equation}
where $T_{amp}$ is the LNA effective noise temperature.

As follows from (2) and (3), the thermal noise peak disappears at $T_{amp} = T_{res}$. This is not a problem when dealing with the room temperature resonators, in which case $T_{res} \sim$ 300 K and $T_{amp} \sim$ 50 K. For that reason, the noise spectrum calculated by Eq. 1 (which was derived assuming $T_{amp} < < T_{res}$) is in reasonable agreement with the measured one. However, in cryogenic experiments, two temperatures can be quite close to each other, i.e., $T_{amp}\approx T_{res}$. This complicates thermal noise detection, as will be shown below.

Here, we describe the use of the dual-channel readout to detect thermal fluctuations emitted by the cryogenic resonator. The latter, along with the LNA attached to its coupling port, was placed at the bottom of the pulse tube cryostat and cooled to 6 K. Fig.~5 shows the photo of the experimental setup. A detailed description of the cryogenic resonator used in our measurements is given in \cite{ivanov2021noise,ivanov2023power}. 

\begin{figure}
    \centering
    \includegraphics[width=0.5\linewidth]{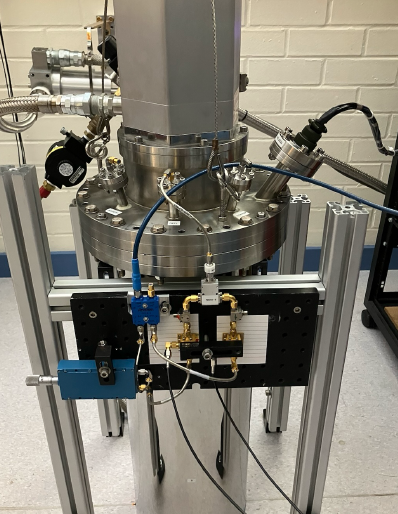}
    \caption{Experimental setup for measuring the thermal noise of the cryogenic resonator. The cryostat houses a pulse-tube cryocooler. The resonator-amplifier assembly is attached to the ``cold finger'' of the pulse tube at the bottom of the cryostat and kept at 6.2 K. The metal plate in the foreground features two power dividers, a mechanical phase shifter, and two double-balanced mixers. A microwave synthesizer driving the mixer's LO port is connected to the setup with a flexible coaxial cable.}
    \label{fig:5}
\end{figure}

The noise measurements were performed at a frequency of approximately 11.2 GHz, with the resonator-synthesizer frequency offset $\Delta F$ equal to 100 kHz. 
Fig.~6 shows the measured rms voltage noise spectra as a function of frequency detuning from the $\Delta F$. The top trace in Fig. 6 corresponds to the single-channel voltage noise spectrum $\delta u_{1}$; the bottom trace is the cross-voltage noise spectrum $\delta u_{1,2}$.
The measurement duration was approximately 2.5 hours, during which $2^{16}$ averages were taken. 
The observed 16 dB decrease in the voltage noise floor (which is 8 dB less than expected for two independent noise processes) is consistent with what was observed earlier with the 50 $\Omega$ terminated LNA (see Fig. 2) and must be caused by the same reasons.
\begin{figure}
    \centering
    \includegraphics[width=1\linewidth]{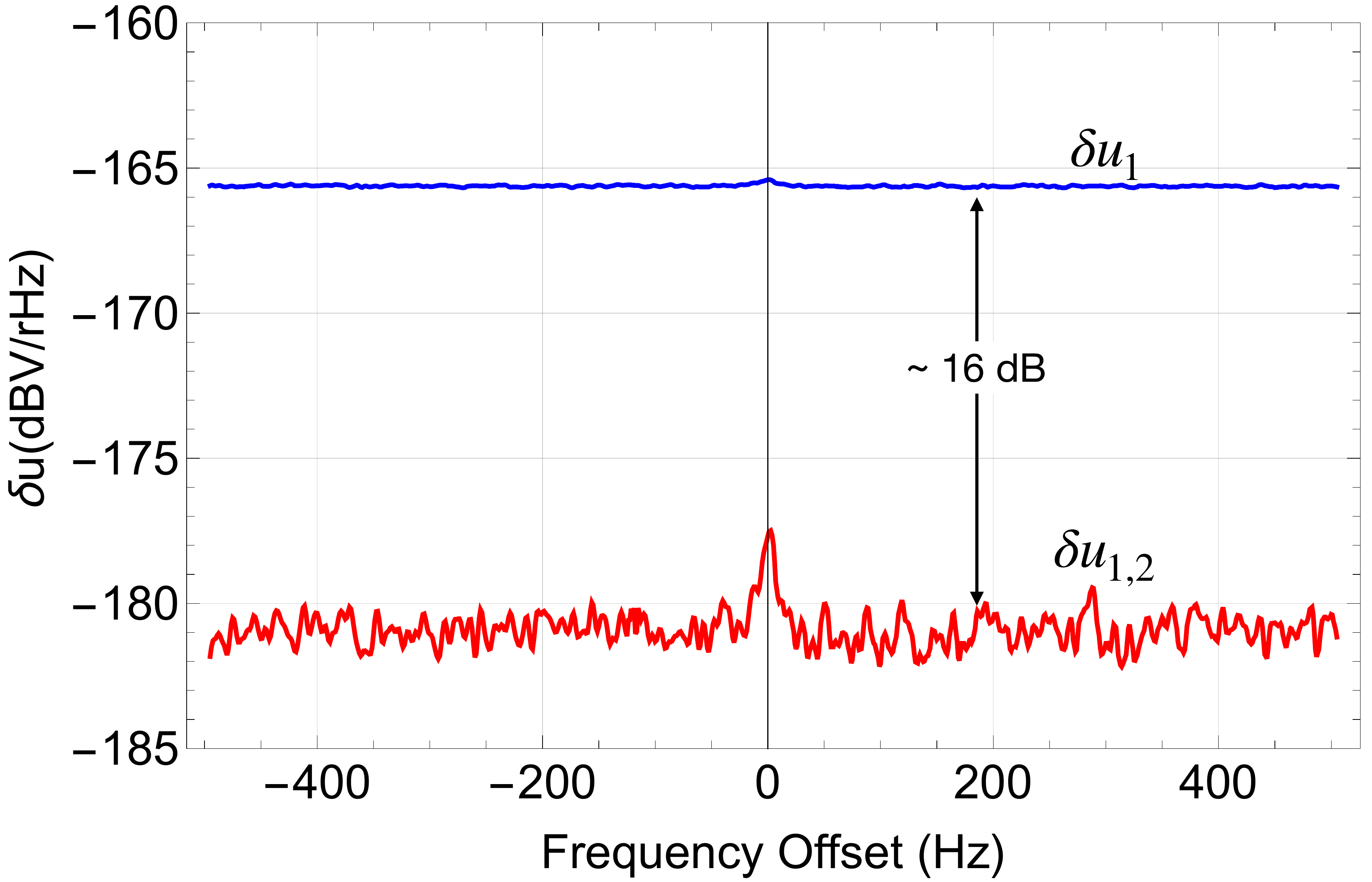}
    \caption{Rms voltage fluctuations at the outputs of the individual channel (top trace), and cross-voltage rms noise (bottom trace).}
    \label{fig:6}
\end{figure}
The noise spectra in Fig.~6 were measured with a Low Noise Factory amplifier with an effective noise temperature of approximately 4 K. Statistical analysis of the data in Fig.~6 shows that the contrasts of the thermal noise peaks, $C_{peak}$, are respectively $1.4\times 10^{-18}$ and $1.0 \times 10^{-18}$ $V_{rms}^{2}$⁄$Hz$ for the top and bottom traces. Furthermore, the standard deviations of the voltage noise spectra, $\sigma_{PSD}$, are approximately $2.0 \times 10^{-20}$ and $1.0 \times 10^{-20}$ $V_{rms}^{2}$⁄$Hz$ for the single and dual-channel measurements. One can consider the ratio $C_{peak}$/$\sigma_{PSD}$ as a figure of merit for comparing the spectral resolutions of the single and dual-channel readouts. Assuming such criteria, the dual-channel readout offers a 1.4-fold improvement (a near 2-fold improvement is expected theoretically \cite{rubiola2010cross}) in the resolution of spectral measurements over its single-channel counterpart.

Interestingly, we could still detect the presence of the cryogenic resonator even when the LNA was placed outside the cryostat. The corresponding voltage noise spectra are shown in Fig.~7. Here, the cross-voltage noise spectrum features a peak with a contrast of 1.6 dB. On the other hand, the spectrum at the output of the single-channel readout exhibits an absorption dip with a depth of $\sim$ 0.3 dB. It is due to the LNA technical fluctuations being absorbed in the cold resonator.

\begin{figure}
    \centering
    \includegraphics[width=1\linewidth]{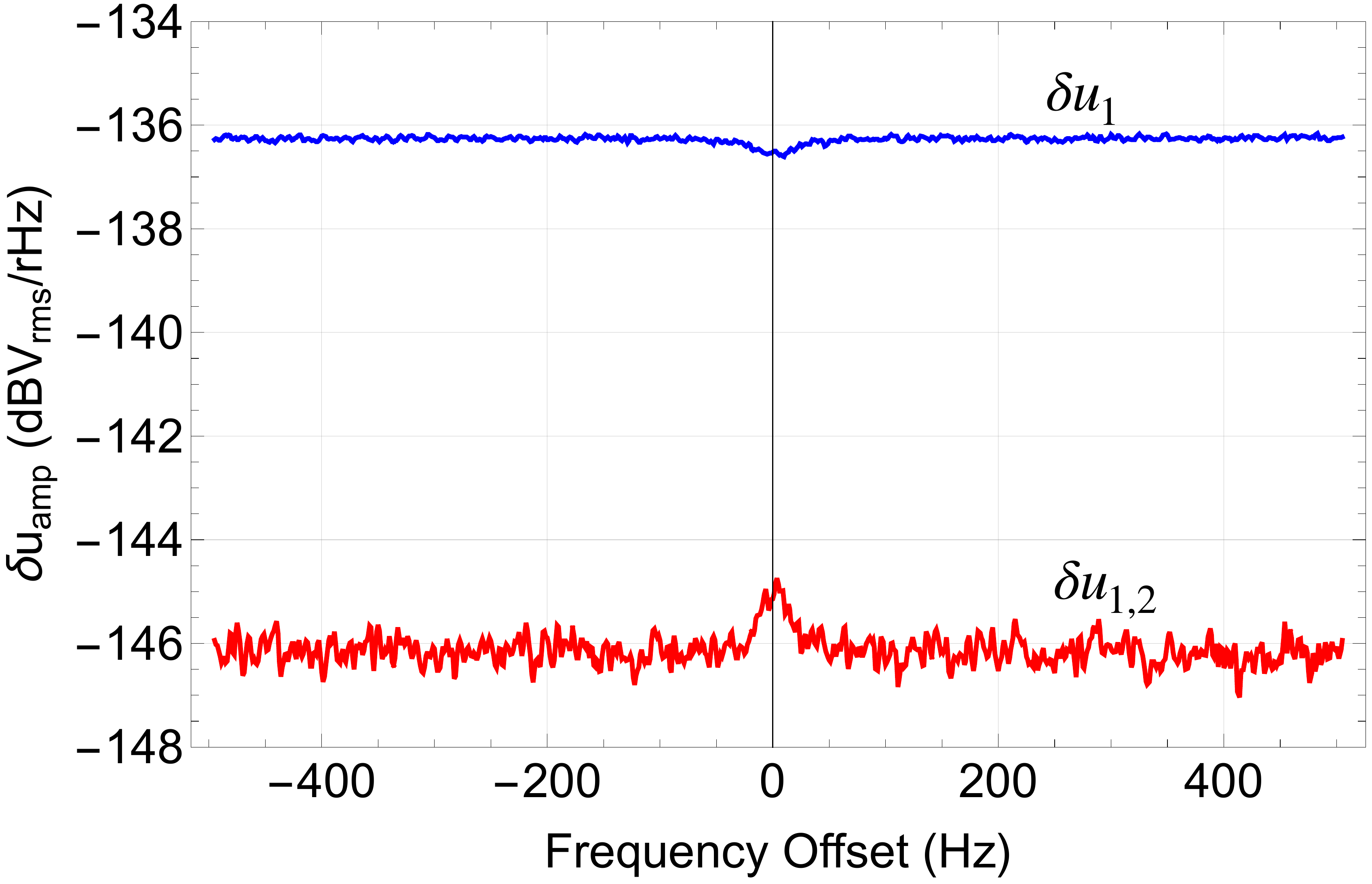}
    \caption{Rms voltage fluctuations at the outputs of the individual channel (top trace), and cross-voltage rms noise (bottom trace): LNA is outside the cryostat}
    \label{fig:7}
\end{figure}

At low frequencies, a quartz tuning fork operating at $2^{15}$ Hz and having a Q-factor of $10^{5}$ can be considered an ideal candidate for measuring its thermal noise output. We made these measurements using an experimental setup shown in Fig.~8(a). As an LNA, we used a non-inverting amplifier with input resistance of 10 M$\Omega$ and capacitance of $\sim$ 25 pF. The amplifier was based on a JFET microchip (OPA656) to minimize the LNA current noise. The latter is a primary noise source that limits the resolution of thermal noise measurements at low frequencies. As shown in \cite{win1raq}, the spectral density of LNAs increases with frequency making the thermal noise measurements in RF quartz resonators, such as the extremely high-Q Boîtiers à Vieillissement Amélioré (BVA) operating at 5 MHz, practically impossible \cite{goryachev2014observation}.

\begin{figure}[htbp]
    \centering
    \begin{subfigure}[b]{0.7\linewidth}
        \includegraphics[width=\linewidth]{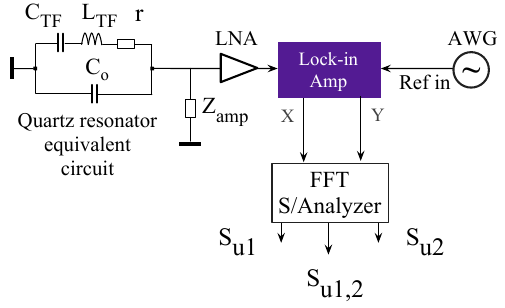}
        \caption{}
        \label{fig:fig8a}
    \end{subfigure}
    \hfill
    \begin{subfigure}[b]{1\linewidth}
        \includegraphics[width=\linewidth]{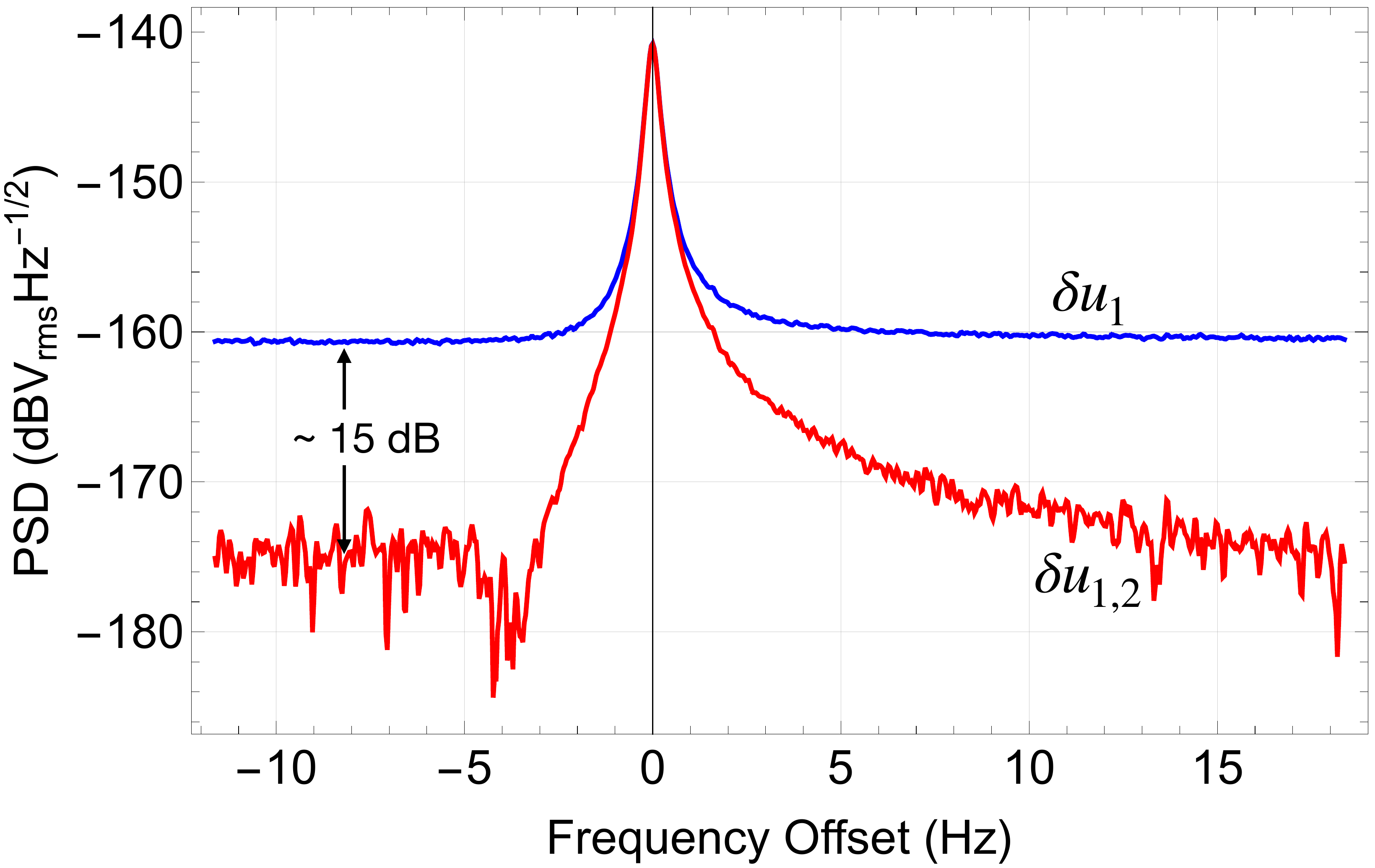}
        \caption{}
        \label{fig:fig8b}
    \end{subfigure}
    \caption{Measurement system (a) and voltage noise spectra at the lock-in amplifier outputs (b). An Arbitrary Waveform Generator (AWG) synchronising the lock-in amplifier is 15 Hz detuned from the tuning fork resonance.}
    \label{fig:8}
\end{figure}

Fig.~8(b) shows the voltage noise spectra at the lock-in amplifier X port (top trace) and the cross-voltage spectrum between X and Y ports (bottom trace). The improvement in the thermal noise peak contrast (when switching from the single to dual-channel measurements) is approximately 16 dB, i.e., similar to what was observed with the microwave resonators. 
For both noise spectra in Fig.~8, as numerical simulations show, the pedestal of the thermal noise peak is due to LNA current fluctuations. The peak in the noise spectra in Fig.~8(b) occurs at frequency 
\begin{equation}
    \omega_{peak} \simeq 1/\sqrt{L_{TF} C_{amp}},
\end{equation}
where $L_{TF}$ is the equivalent inductance of the tuning fork, and $C_{amp}$ is the LNA input capacitance. A ``dent'' in the cross-voltage noise spectrum a few Hz below the thermal noise peak is at the resonant frequency of the tuning fork
\begin{equation}
    \omega_{res}\simeq 1/\sqrt{L_{TF} C_{TF}}.
\end{equation}
At this frequency, the impedance of the quartz tuning fork is approximately 20 k$\Omega$, as compared to $\sim$ 10 M$\Omega$ at the $\omega_{peak}$.

The experiments with the dual-channel noise measurement system demonstrated that it could suppress the broad-band fluctuations common to both channels.  This proved possible due to the combined use of the IQ-mixing and cross-correlation signal processing. When detecting the thermal noise exhibited by microwave or mechanical electromagnetic resonators, we observed a 15 to 17 dB improvement in the thermal noise peak contrast. Additionally, in experiments with cryogenic microwave resonators, we found that the dual-channel readout system provided a 1.4-fold improvement in the resolution of spectral measurements. The newly discovered ability of the dual-channel readout system to discriminate between broad- and narrow-band fluctuations may benefit the laboratory search for dark matter, especially an approach which involves the use of cryogenic microwave resonators.

\begin{acknowledgments}
This work was funded by the ARC Centre of Excellence for Engineered Quantum Systems, CE170100009, Dark Matter Particle Physics, CE200100008, and by the Defence Science Centre, an initiative of the State Government of Western Australia.
\end{acknowledgments}

\bibliographystyle{aipnum4-2}
\bibliography{cross.bib}

\end{document}